\documentstyle[12pt]{article}
\begin{document}

\title{Quantum computation by quantum-like systems}
\author{M. A. Man'ko\thanks{%
P. N. Lebedev Physical Institute, Leninskii Prospect 53, Moscow 117924,
Russia, e-mail: mmanko@sci.lebedev.ru} \thanks{%
Zentrum f\"{u}r interdisziplin\"{a}re Forschung, Universit\"{a}t Bielefeld,
Wellenberg 1, 33615 Bielefeld, Germany} , V. I. Man'ko\footnotemark[1]  
\footnotemark[2]  and R. Vilela Mendes\thanks{%
Grupo de F\'{\i }sica Matem\'{a}tica, Complexo Interdisciplinar,
Universidade de Lisboa, Av. Gama Pinto, 2 - P1699 Lisboa Codex, Portugal,
e-mail: vilela@cii.fc.ul.pt} \footnotemark[2] }
\date{}
\maketitle

\begin{abstract}
Using a quantumlike description for light propagation in nonhomogeneous
optical fibers, quantum information processing can be implemented by optical
means. Quantum-like bits (qulbits) are associated to light modes in the
optical fiber and quantum gates to segments of the fiber providing an
unitary transformation of the mode structure along a space direction.
Simulation of nonlinear quantum effects is also discussed.
\end{abstract}

PACS:\ 03.67.Lx

\section{Introduction}

The quantum computer idea~\cite{Ben,Feyn,Deut} uses the possibility to code
numerical information by vectors in Hilbert space. In the simplest case, a
two-dimensional Hilbert space is said to code a {\it qubit}. A physical
realization of a qubit might be a spin 1/2 particle or a two-level atom.
Calculations are carried out by unitary time-evolution transforming an input
state vector into a final state vector in Hilbert space. Several
characteristic features distinguish the quantum computation paradigm from
classical computation. First, because of the superposition principle, the
qubit space contains all the complex linear combinations, rather than just
two states as in a classical bit. Second, in a quantum space of $n$ qubits
(of dimensionality $2^{n}$) there are both factorized and entangled states,
the latter having no correspondence in the space of $n$ classical bits.
Third, quantum evolution operates simultaneously in the $2^{n}$ states,
implying intrinsic exponential parallelism of quantum computation. To
extract the result from a quantum computation one has to observe the system,
that is, to project it in one of the exponentially many states, thus losing
most of the exponential amount of information generated by the unitary
evolution. However, it is possible to take advantage of the exponentiality
of quantum computation using the interference mechanism characteristic of
quantum mechanics. In short, it is the combined effect of {\it entanglement}%
, {\it exponential parallelism} and {\it interference} that may allow the
full potential of quantum computation to be realized \cite{Aharonov}.

According to the Church thesis, a classical Turing machine can simulate the
computation of any computable function. Therefore a classical computer can
simulate any computation of a quantum computer. The problem is how long the
simulation will take to run. In particular it is know that some problems
that are computed in exponential time by classical computers might be solved
in polynomial time by quantum computers. Therefore, to take full advantage
of the quantum computation algorithms, one requires physical devices obeying
quantum laws.

There are many classical systems that physically implement some of the
features of quantum computation\cite{Spree,FP,CerfPR,Kwiat}. For example
electromagnetic waves may be linearly superimposed and interfere.
Nevertheless, given the qualitative differences between classical and
quantum mechanics, no classical system, {\it where computations correspond
to evolution in real time}, may ever implement simultaneously all the
features of quantum computation. Otherwise we would have proved the physical
equivalence of classical and quantum mechanics.

Notice however that quantum computation {\it is not} quantum mechanics.
Quantum computation is a mathematical algorithm that uses all the
mathematical features of quantum mechanics. In particular it is irrelevant
for the algorithm how the Hilbert space is physically implemented and
whether the unitary evolution is taking place along a real time direction or
along some other coordinate. It is here that quantum-like systems may play a
role. Quantum-like systems are classical systems which obey equations
formally identical to the Schr\"{o}dinger equation, but where the role of
time is played by a space coordinate. Therefore, insofar as they obey
equations mathematically identical to those of quantum mechanics, they may
implement all the quantum computation operations, provided the unitary
evolution is interpreted not as evolution in time, but as evolution along a
space coordinate. Because of this exchange of the role of the coordinates,
there is no contradiction with the non-equivalence of classical and quantum
mechanics.

In Sect.2 we give an overview of quantum-like systems and then,
concentrating on fiber optics phenomena, we discuss how quantum information
may be coded on the fiber and the kind of non-homogeneities and interactions
needed to implement a set of universal quantum gates. Here we have
concentrated on harmonic light modes. An alternative scheme might be
developed based on soliton propagation on the fibers.

The fact that the unitary evolution needed for a quantum gate is obtained by
setting up a nonuniform refraction index profile along the fiber, means that
the unitary evolution becomes permanently coded in the hardware. This may be
physically more convenient than to control a sequence of operations each
time the gate is operated.

Abrams and Lloyd\cite{Abrams} have suggested that a stronger computation
model would be obtained with non-linear terms in the quantum evolution. So
far, all experimental evidence favors strict linearity of quantum mechanics.
However, this is not so in quantum-like systems where non-linear effects may
easily be introduced. Actual implementations of non-linear quantum
computation might therefore use quantum-like systems.

\section{Quantum-like systems}

More than half a century ago, Fock and Leontovich have shown that paraxial
beams of electromagnetic radiation in the parabolic approximation can be
described by a Schr\"{o}dinger-like equation~\cite{Leon,Fock-Leon}. The role
of time in this equation is played by the spatial (longitudinal) coordinate
of the light beam, the role of Planck's constant is played by the light
wavelength, and the role of potential energy by the index of refraction of
the medium. Thus, the paraxial beam of light, a purely classical object, is
a quantum-like system obeying equations formally identical to those of
quantum mechanics.

Given the Helmholz equation for a component of the electric field, obtained
for a fixed frequency, neglecting media dispersion and polarization 
\begin{equation}
\frac{\partial ^{2}E}{\partial ^{2}x}+\frac{\partial ^{2}E}{\partial z^{2}}%
+k^{2}n^{2}(x,z)\,E=0  \label{5}
\end{equation}
(we use the planar configuration, $\lambda =2\pi /k$ is the wavelength in
vacuum, and $z$ the longitudinal coordinate).

Introduce the complex function $\psi (x,z)$, which is the slowly varying
amplitude of the electric field in 
\begin{equation}
E(x,z)=n_{0}^{-1/2}(z)\,\psi (x,z)\,\exp \left[ ik\int_{0}^{z}n_{0}(\xi
)~d\xi \right]  \label{4}
\end{equation}
The ansatz~(\ref{4}) reduces the Helmholz equation~(\ref{5}) to a
Schr\"{o}dinger-like equation 
\begin{equation}
i\lambda \,\frac{\partial \psi (x,z)}{\partial z}=-\frac{\lambda ^{2}}{4\pi
n_{0}(z)}\,\frac{\partial ^{2}\psi (x,z)}{\partial x^{2}}+U(x,z)\,\psi (x,z)
\label{3}
\end{equation}
$U(x,z)$ being an effective potential related to the index of refraction of
the medium, $n(x,z)$%
\[
U(x,z)=\frac{\pi }{n_{0}(z)}\,\left[ n_{0}^{2}(z)-n^{2}(x,z)\right] \, 
\]
and $n_{0}(z)=n(0,z)$ the index of refraction at the beam axis. In deriving
this equation, second-order $z-$derivatives of $\psi $ and derivatives of
the function $n_{0}(z)$ were neglected. This is justified for slow variation
of the index of refraction along the beam axis over distances of order of
one wavelength 
\[
\frac{\lambda }{n_{0}^{2}(z)}\left| \frac{dn_{0}(z)}{dz}\right| \ll 1\,. 
\]

The Fock--Leontovich approximation is a basis for the description of
light-beam propagation in optical fibers~\cite
{Gloge-Marcuse,Marcuse,Arnaud,MankoLeeSpr} leading to 
\begin{eqnarray}
i\lambda \,\frac{\partial \psi (x,y,z)}{\partial z} &=&-\frac{\lambda ^{2}}{%
4\pi n_{0}(z)}\left( \frac{\partial ^{2}\psi (x,y,z)}{\partial x^{2}}+\frac{%
\partial ^{2}\psi (x,y,z)}{\partial y^{2}}\right)  \label{6} \\
&&+\frac{\pi }{n_{0}(z)}\left[ n_{0}^{2}(z)-n^{2}(x,y,z)\right] \psi (x,y,z)
\nonumber
\end{eqnarray}
which is Schr\"{o}dinger-like for a wave function depending on the
transversal coordinates $x$ and $y$. The longitudinal $z$ coordinate plays
the role of time in the Schr\"{o}dinger equation. The unitary $z$-evolution
of the electromagnetic complex amplitude is described by the evolution
operator $\hat{U}(z)$ 
\[
\hat{U}(z,z_{0})\psi (x,y,z_{0})=\psi (x,y,z), 
\]
associated to the Hamiltonian 
\begin{equation}
\hat{H}(z)=\left( \frac{\hat{p}_{x}^{2}}{2}+\frac{\hat{p}_{y}^{2}}{2}\right) 
\frac{1}{n_{0}(z)}+U(x,y,z).  \label{7}
\end{equation}
with $\hat{p}_{x}=-i\lambda \,\frac{\partial }{\partial x}\,,\hat{p}%
_{y}=-i\lambda \,\frac{\partial }{\partial y}\,$ and a potential function 
\[
U(x,y,z)=\frac{\pi }{n_{0}(z)}\left[ n_{0}^{2}(z)-n^{2}(x,y,z)\right] . 
\]

Other quantum-like systems are reviewed in~\cite{Erici,Caserta}. An
important example is sound-wave propagation in acoustic waveguides~\cite
{Brekhovskikhbook}. Acoustic waves in the paraxial approximation are well
described by a Schr\"{o}dinger-like equation. Charged-particle beams were
also recently discussed as quantumlike systems~\cite{FedMie,FedManPRE}.
Light beams inside diode lasers have been treated as quantumlike systems as
well, due to the waveguide structure of their active region~\cite
{Gevork,Rita}. As remarked in \cite{Bielefeld} the variety of quantumlike
classical systems provides a wide range of possibilities for perfect
simulation of quantum computation operations by classical means.

\section{Modes and gates in optical fibers}

In an optical fiber, the light modes, being solutions of the
Schr\"{o}dinger-like equation (\ref{6}), have all the properties of
quantum-mechanical wave functions including the entanglement phenomenon.

Light modes in the fiber are used to code quantum-like bits ({\it qulbits}).
How many qulbits may live in one optical fiber? In the simplest case, which
is considered here, light modes with a fixed frequency are considered. But,
in the same optical fiber, light of different frequencies may be used. The
Helmholtz equation holds for each frequency, with an index of refraction
profile that may be different for different frequencies. Hence, in the same
optical fiber, one may store many qulbits simply by exploiting the
propagation of light beams with different frequencies. Interaction of
photons with different frequencies may provide useful computation effects.

The Fock--Leontovich approximation is obtained from the Helmholtz equation
for the components of the electric or magnetic field. This equation is a
scalar approximation which neglects the tensorial structure of the
dielectric constant and the polarization. Also neglected are time and space
dispersion, related to the nonlocal linear response of the medium to
electromagnetic perturbations. Taking into account these effects would
provide an even richer framework to accommodate qulbits in the fiber.

We consider now the scalar fixed frequency situation described by Eq.~(\ref
{6}). When the index of refraction profile has the form of an inverse well,
the optical fiber traps discrete modes $\psi _{n_{1}n_{2}}(x,y,z)$, $n_{1}$
and $n_{2}$ being integer labels.

With light modes on a fiber and the unitary $z-$evolution associated to Eq.(%
\ref{6}) one may perform quantum computation over continuous variables in a
way similar to the one proposed for time evolution in Ref.\cite{Lloyd1}. All
the physical interactions needed for the construction of polynomial
Hamiltonians are available by the choice of the appropriate refraction
profile and by Kerr interactions. Also, as explained below, by restricting
oneself to finite-dimensional subspaces of excitations, one may perform the
same operations as in quantum computation with discrete variables.

There are several ways to code information by light modes in a fiber which
may be useful for quantum information processing, in particular those
associated to different choices of basis in self-focusing potentials. The
important self-focusing case is associated to quadratic potentials of the
form 
\begin{equation}
U(x,y,z)=a(z)x^{2}+b(z)y^{2}+d(z)xy+e(z)x+f(x)y+l(z),  \label{3.1}
\end{equation}
and in this case, an explicit solution may be obtained for the four $z-$%
dependent integrals of motion\cite{MankoLeeSpr} 
\begin{equation}
\left( 
\begin{array}{l}
\hat{x}_{0}(z) \\ 
\hat{y}_{0}(z)
\end{array}
\right) =\hat{U}(z,z_{0})\left( 
\begin{array}{l}
\hat{x} \\ 
\hat{y}
\end{array}
\right) \hat{U}^{-1}(z,z_{0}),\qquad \left( 
\begin{array}{l}
\hat{p}_{x0}(z) \\ 
\hat{p}_{y0}(z)
\end{array}
\right) =\hat{U}(z,z_{0})\left( 
\begin{array}{l}
\hat{p}_{x} \\ 
\hat{p}_{y}
\end{array}
\right) \hat{U}^{-1}(z,z_{0})  \label{3.2}
\end{equation}
Defining the boson integrals of motion 
\begin{equation}
\left( 
\begin{array}{l}
a(z) \\ 
a^{\dagger }(z)
\end{array}
\right) =\frac{1}{\sqrt{2}}\left( 
\begin{array}{l}
\hat{x}_{0}(z)+i\hat{p}_{x0}(z) \\ 
\hat{x}_{0}(z)-i\hat{p}_{x0}(z)
\end{array}
\right) ,\left( 
\begin{array}{l}
b(z) \\ 
b^{\dagger }(z)
\end{array}
\right) =\frac{1}{\sqrt{2}}\left( 
\begin{array}{l}
\hat{y}_{0}(z)+i\hat{p}_{y0}(z) \\ 
\hat{y}_{0}(z)-i\hat{p}_{y0}(z)
\end{array}
\right)  \label{3.3}
\end{equation}
several basis may be constructed. The discrete Fock-state modes are
solutions to the eigenvalue equation 
\begin{eqnarray}
a^{\dagger }(z)a(z) &\mid &n_{1},n_{2},z\rangle =n_{1}\mid
n_{1},n_{2},z\rangle ,  \label{S9} \\
b^{\dagger }(z)b(z) &\mid &n_{1},n_{2},z\rangle =n_{2}\mid
n_{1},n_{2},z\rangle ,\qquad n_{1},n_{2}=0,1,2,\ldots  \label{S10}
\end{eqnarray}
The Fock-state modes are obtained from the fundamental mode $\mid
0,0,z\rangle $ by 
\begin{equation}
\mid n_{1},n_{2},z\rangle =\frac{a_{1}^{\dagger n_{1}}(z)b^{\dagger n_{2}}(z)%
}{\sqrt{n_{1}!n_{2}!}}\mid 0,0,z\rangle .  \label{S24}
\end{equation}

A spin-like description of the Fock-state modes is related to $SU(2)$%
-subgroup of the Weyl-symplectic group in two dimensions. This is related to
the Jordan--Schwinger map 
\begin{equation}
J_{+}(z)=a^{\dagger }(z)b(z);\quad J_{-}(z)=b^{\dagger }(z)a(z);\quad
J_{3}(z)=\frac{1}{2}\left( a^{\dagger }(z)a(z)-b^{\dagger }(z)b(z)\right)
\label{3.4}
\end{equation}
Irreducible representation spaces for this subgroup are spanned by states $%
\left\{ \mid n_{1},n_{2},z\rangle ;n_{1}+n_{2}=N\right\} $ for each fixed $N$%
.

On the other hand, coherent modes in the optical fiber are labelled by two
complex numbers $\alpha $ and $\beta $, 
\begin{equation}
\mid \alpha ,\beta ,z\rangle =\exp \left[ \alpha a^{\dagger }(z)-\alpha
^{*}a(z)\right] \exp \left[ \beta b^{\dagger }(z)-\beta ^{*}b(z)\right] \mid
0,0,z\rangle ,  \label{S26}
\end{equation}
Using this basis, the self-focusing fiber could be considered a Gaussian
channel for numerical information. For arbitrary choices of the complex
numbers $\alpha $ and $\beta $ this is an overcomplete set. However choosing
the numbers on the von Neumann lattice\cite{Boon} 
\begin{equation}
\alpha _{m_{1},n_{1}}=\frac{1}{\sqrt{2}}\left( n_{1}+i2\pi m_{1}\right)
;\quad \beta _{m_{2},n_{2}}=\frac{1}{\sqrt{2}}\left( n_{2}+i2\pi m_{2}\right)
\label{3.5}
\end{equation}
and excluding two pairs (for example $m_{1}=n_{1}=m_{2}=n_{2}=0$) one
obtains a discrete {\it complete} set of coherent modes, which might provide
a basis for the coding of a large amount of quantum-like information in the
fiber.

We now analyze the question of what unitary transformations may be obtained
by evolution of the light modes along the fiber. The simplest possibility is
by a change of the index of refraction profile. From Eq.(\ref{6}) we may
write a path integral representation for the evolution of the light mode
along the fiber 
\begin{equation}
\psi (x,y,z)=G\left( x,x_{0},y,y_{0},z\right) \psi (x_{0},y_{0},z)
\label{N1}
\end{equation}
with 
\begin{equation}
G\left( x,x_{0},y,y_{0},z\right) =\int_{\left( x_{0},y_{0},0\right)
}^{\left( x,y,z\right) }d^{2}\overrightarrow{\xi }\exp \left\{ \frac{i}{%
\lambda }\int_{0}^{z}d\tau \left[ \pi n_{0}(z)\stackrel{\bullet }{\xi }%
^{2}(z)-U(\overrightarrow{\xi },z)\right] \right\}  \label{N2}
\end{equation}
$\overrightarrow{\xi }$ being a two-dimensional vector on the fiber sections
and 
\begin{equation}
U(\overrightarrow{\xi },z)=\frac{\pi }{n_{0}\left( z\right) }\left[
n_{0}^{2}(z)-n^{2}(\overrightarrow{\xi },z)\right]  \label{N3}
\end{equation}
One sees from Eq.(\ref{N2}) that adjusting the index of refraction profile
changes not only the potential but also the coefficient of the kinetic term.
Let us consider a self-focusing quadratic potential and define, as before, $%
a^{\dagger },a$ and $b^{\dagger },b$ to be creation and annihilation
operators for harmonic modes along the $x$ and $y-$directions respectively.
For simplicity we drop the $z$ argument in the operators. Then, changing the
index of refraction profile gives us direct access to the generators 
\begin{equation}
a^{\dagger }a+b^{\dagger }b;a^{\dagger }+a;\left( a^{\dagger }+a\right)
^{2};b^{\dagger }+b;\left( b^{\dagger }+b\right) ^{2};\left( a^{\dagger
}+a\right) \left( b^{\dagger }+b\right)  \label{N4}
\end{equation}
By a simple reasoning using the Baker-Campbell-Hausdorff formula one
concludes that by a non-uniform change of the index of refraction one may
obtain in Eq.(\ref{N2}) all the operations of the Weyl-symplectic group in
two dimensions. This group is not compact. Therefore the unitary
representations of the full group are not finite-dimensional. This might be
explored for information manipulation schemes were an unbounded number of
states is manipulated. However for operations on a finite number of qulbits,
it is the compact subgroups that are important. In particular a useful
subgroup is the $SU\left( 2\right) $ group described before in Eq.(\ref{3.4}%
). Finite-dimensional irreducible spaces for this subgroup are 
\begin{equation}
a^{\dagger n}b^{\dagger m}|0>,\quad n+m=2k  \label{N6}
\end{equation}
for $k=0,\frac{1}{2},1,\frac{3}{2},2,...$ with dimension $2k+1$. In this
finite-dimensional spaces all unitary operations may be implemented on the
fiber by changing the index of refraction profile.

To perform universal quantum computation it is necessary, at least, to have
arbitrary unitary transformations on a single qulbit and a CNOT operation on
two qulbits\cite{Barenco}. According to the discussion above the $\left|
0\right] ,\left| 1\right] $ qulbit states may be coded, for example, as
follows 
\begin{equation}
\left| 1\right] =a^{\dagger }|0>,\quad \left| 0\right] =b^{\dagger }|0>
\label{N7}
\end{equation}
This being a $k=\frac{1}{2}$ two-dimensional representation, all unitary
transformations may be performed in this space by the $SU\left( 2\right) $
subgroup.

For the CNOT\ operation we may code the $\left| 1\right] -$state of the
control bit as the application of two energy quanta along the $x-$ direction
and the $\left| 0\right] -$state of the control bit as the application of
two energy quanta along the $y-$direction. For the target bit we use the
same coding as in (\ref{N7}). Therefore 
\begin{equation}
\left| 11\right] =a^{\dagger 3}|0>,\left| 01\right] =a^{\dagger 2}b^{\dagger
}|0>,\left| 10\right] =a^{\dagger }b^{\dagger 2}|0>,\left| 00\right]
=b^{\dagger 3}|0>  \label{N8}
\end{equation}
the first label in $\left| \alpha \beta \right] $ being the label of the
target qulbit and the second the label of the control qulbit. The subspace
spanned by (\ref{N8}) is a four-dimensional $SU(2)-$irreducible subspace.
Therefore all unitary transformations may be implemented in this subspace
and, in particular, the CNOT operation.

We have therefore proved that, with this coding, the self-focusing fiber is
capable of universal quantum computation. With higher order potentials many
other possibilities would be available. For example, an interacting term of
the form $\eta a^{\dagger }ab^{\dagger }b$ appears in the Kerr Hamiltonian.
This allows to perform phase operations on $a^{\dagger }$ modes gated by the 
$b^{\dagger }$ excitations. For example by coding the target qulbit as 
\[
\left| 1\right] =\frac{1}{\sqrt{2}}\left( 1-a^{\dagger }\right) |0>,\quad
\left| 0\right] =\frac{1}{\sqrt{2}}\left( 1+a^{\dagger }\right) |0> 
\]
and the control qulbit as 
\[
\left| 1\right] =b^{\dagger }|0>,\quad \left| 0\right] =|0> 
\]
the operator $\exp \left( i\pi a^{\dagger }ab^{\dagger }b\right) $
implements the CNOT gate.

The self-focusing fiber is a versatile medium to code quantum-like
information and this is the reason we have emphasized the existence of
several light mode basis. In Eq.(\ref{N7}) above, a qulbit is coded using
modes in the $x$ and $y$ directions. We might as well have used $x-$modes
only and coded the qulbit using the quadratures $\hat{x}$ and $\hat{p}_{x}$.
Then, light propagation with the symmetric self-focusing Hamiltonian
corresponding to the operator 
\[
U=e^{i\left[ (\hat{p}_{x}^{2}/2+(\hat{x}^{2}/2)\right] \pi /4} 
\]
yields 
\[
\left( 
\begin{array}{c}
\hat{p}_{x} \\ 
\hat{x}
\end{array}
\right) \rightarrow \frac{1}{\sqrt{2}}\left( 
\begin{array}{c}
\hat{p}_{x}+\hat{x} \\ 
\hat{p}_{x}-\hat{x}
\end{array}
\right) , 
\]
a Hadamard gate transformation. Observation of such transformation can be
done measuring the position and direction of rays in the optical fiber for
Gaussian wave packets.

\section{Conclusions}

1) Classical systems that are quantum-like, in the sense that their
evolution along a space direction is described by a Schr\"{o}dinger
equation, possess a high potential for information processing including
quantum computation. A promising system of this type consists of a light
beam propagating along an optical fiber. It should be noticed however that
similar possibilities exist with other systems, for example acoustic waves
propagating along an acoustic waveguide. Practical implementation problems
to be addressed are the choice of the optical fiber and the definition of a
coding standard for the qulbits. The variety of materials used and a fairly
well developed optical fiber technology give us hope that the model
Hamiltonians needed for the operations of quantum computing may be
physically implemented in this medium.

2) The fact that the unitary evolution of the quantum-like systems is
associated to a space dimension, means that the unitary transformation is
implemented in the hardware, rather than requiring a precise sequence of
temporal operations. Also and this might be useful for mass production
purposes, once a non-uniform refraction index profile is set up on the
material, many different gates may be obtained simply by cutting fiber
segments of different lengths.

3) Finally we should also point out the easy possibility to imitate
nonlinear quantum mechanics by means of the nonlinear media response to the
electromagnetic radiation. This might provide, as suggested in \cite{Abrams}%
, even more powerful quantum computation algorithms.

\end{document}